\documentclass[10pt,letterpaper]{article}
\usepackage{opex3}

\newcommand{\tW}{t_{\rm{W}}}
\newcommand{\tauW}{\tau_{\rm{W}}}

\newcommand{\kL}{k_{\rm{L}}}

\newcommand{\zT}{z_{\rm{T}}}
\newcommand{\ld}{l_{\rm{d}}}
\newcommand{\losc}{l_{\rm{osc}}}

\newcommand{\chin}{\vert \chi_n \rangle}
\newcommand{\phim}{\vert \phi_m \rangle}

\newcommand{\nex}{n_{\rm{e}}}
\newcommand{\nor}{n_{\rm{o}}}

\newcommand{\Ieff}{I_{\rm{eff}}}
\newcommand{\Isat}{I_{\rm{sat}}}

\newcommand{\IW}{I_{\rm{W}}}
\newcommand{\Imax}{I_{\rm{max}}}
\newcommand{\Iavg}{I_{\rm{avg}}}

\newcommand{\dnmax}{\Delta n _{\rm{max}} }

\newcommand{\gradp}{\nabla^2_{\perp}}

\newcommand{\dnmaxt}{\Delta n_{\rm{max}}^{\rm{theo}}}
\newcommand{\dnt}{\Delta n^{\rm{theo}}}

\begin{document}

%



\title{Absolute calibration of the refractive index in photo-induced photonic lattices}
\author{Julien Armijo$^{1, *}$, Rapha\"el Allio$^{1,2}$ and Cristian Mej\'ia-Cort\'es$^1$}

\address{
$^1$Departamento de Fisica, MSI-Nucleus on Advanced Optics, and Center for Optics and
Photonics (CEFOP), Facultad de Ciencias, Universidad de Chile, Santiago, Chile\\
$^2$ Universit\'e de Rennes I, France}

\email{$^*$ julienarmijo@gmail.com}


\begin{abstract} 
We demonstrate a method to experimentally calibrate the refractive index modulation in photorefractive lattices, a task rarely addressed that is crucial for quantitative comparisons of theories with experiments.
We consider the linear propagation of a normally incident plane wave through simple lattices and its modulation amplitude at crystal output face.
Finding no evidence of longitudinal (Talbot-like) oscillations, we discard an ideal propagation theory and construct a simple effective model that includes longitudinal relaxation.
We obtain calibrations of 1D and 2D lattices consistent with standard theory in a high saturation regime.
For 2D lattices, we find anisotropies $\chi=1.5- 2.5$, stronger for smaller lattice period, and refractive indexes larger than for 1D lattices, also with more noise. 
\end{abstract}



\section{Introduction}

Photorefractive crystals are materials in which refractive index patterns can be induced by illumination with structured light \cite{yeh93}.
This effect is complex, intrinsically anisotropic and nonlocal, featuring various terms and regimes \cite{kukhtarev78, zozulya95}. 
Due to several interesting applications, its has been pursued by many groups.
In the last decades, photo-induced waveguide arrays, or photonic crystals, have been used to study the propagation of linear and non-linear light waves in various lattice structures, allowing the observation of discrete optical solitons \cite{fleischer03}, discrete optical vortices in 2D lattices \cite{neshev04}, or Anderson localization of light in disordered landscapes \cite{schwartz07}, among many others.

However, despite these numerous realizations, an absolute experimental calibration of the lattice strength, i.e., its refractive index amplitude, has, to the best of our knowledge, not been performed systematically for the interesting parameters.
The lattice strength is a crucial parameter for the quantitative comparisons of theoretical predictions with experimental data.
In previous works reporting photorefractive data and simulations, the choice of parameters in simulations is not always justified (see, e.g. \cite{schwartz07}).
Usually the estimation of experimental parameters relies on the complex Kukhtarev theory, but the parameters used for these estimates are also rarely clearly justified.
Some works also mention measured lattice strengths being used for simulations, without specifying the measurement method (see, e.g. \cite{sun12}).

Comparing theories with experiments is a central goal in physical sciences. To do so, it is most desirable to calibrate the key experimental parameters with simple and direct methods, independently form complex theories resting on several approximations and hardly accessible quantities. 
An absolute calibration method should be a method allowing to obtain from some measured quantiti(es), a numerical estimation of the parameter that is not relative to some unmeasured reference value (as in \cite{boguslawski13}), but is an absolute number.
While for cold atoms in optical lattices, efficient calibration methods have been developed for the lattice strength \cite{denschlag02}, the analog calibration for photorefractive lattices is to our knowledge not addressed.
For photonic crystals generated with the femtosecond writing technique, a calibration method based on near-field microscopy of light going out of the crystal, is available \cite{blomer06}. However it can generally not be used for photorefractive lattices, since it requires monomode waveguides to invert the Helmholtz equation \cite{mansour96}.
In the context of photorefractive holographic recording, far-field diffraction efficiencies for plane waves at the Bragg angle have often been measured for calibrating the patterns (see, e.g., \cite{kukhtarev78, magnusson76}), using, e.g.,  the Kogelnik formula \cite{kogelnik69}. However in the more recent context of photonic lattices, such methods have apparently been left aside, perhaps due to different parameter ranges used in this context (thicker crystals, etc.).

For photorefractive lattices, the most common procedure recently to estimate the lattice strength is to rely on the Kukhtarev model \cite{kukhtarev78}, assuming a steady-state. 
The most complete approach is the full anisotropic treatment \cite{zozulya95}, but often further simplifications are used, for example, neglecting the diffusion term in carrier transport, the anisotropy, or the residual nonlinearity affecting the ordinarily polarized waves. 
Even in the most complete model, one should keep in mind that the microscopic processes underlying the photorefractive effect are complex non-equilibrium quantum many-body phenomena. 
The Kukhtarev theory is a highly simplified macroscopic description, which assumes constant phenomenological parameters and kinetic rates (absorption, mobilities, etc.), for which measurements are rarely available.
For example, the saturation intensity (or "dark intensity") $\Isat$, a crucial parameter of the theory \cite{zozulya95}, is introduced heuristically as a constant quantity accounting for thermal carrier generation, but often it is simply assumed equal to the intensity of some background light.
Also, most works consider the photorefractive steady-state, completely neglecting dynamical aspects. 
However, due to its complexity, the photorefrative effect can be non-stationary \cite{zozulya95, zozulya96}, not only in time, but also along propagation in the crystal.
Moreover, it may be useful for experiments (see, e.g., \cite{boguslawski13}), to exploit transient photorefractive writing conditions, for example to access different lattice strengths.
To illustrate the roughness of the standard theory, one can recall, in the original paper \cite{kukhtarev78}, that the model (in the simple purely diffusive case) was found to match observations qualitatively well, but with quantitative discrepancies of order 200-300\%.

In this paper, we develop a direct calibration method, based on the linear propagation of plane waves in photo-induced lattices observed in the real space (near-field).
Firstly, we explain why the often mentioned method of digital holography is inappropriate for the photonic lattices with typically interesting parameters, due to longitudinal quasi-oscillations (LQO) analog to a Talbot effect \cite{berry96, iwanow05}.
We then observe experimentally that an ideal propagation theory is insufficient since it predicts LQO that we do not observe at the crystal output face. 
To interpret our data, we construct a phenomenological model, the simplest that we found able to reproduce both the observed damping of LQO, and the observed saturation of the modulation amplitude $\alpha_1 \simeq 1.1$ at large lattice strength.
Our model has only one heuristic parameter, that can easily be extracted from measurements in a particular experiment.
Due to its simplicity, it is robust and self-consistent, although its accuracy is not expected to be substantially better than 30\%.

For 1D lattices, we estimate the lattice strength at any writing time, and find values consistent with standard theory.
For 2D lattices, we find lattice anisotropies $\chi=1.5- 2.5$, stronger for smaller lattice period, and higher refractive indexes than in 1D lattices. 
Calibrations for 2D lattices are more noisy, 
which probably indicates stronger deformations of the waveguides by non-ideal effects (anisotropy, diffusive mechanism, nonlinearity) that reduce the validity of our approximations.
Our observations illustrate the importance of performing direct calibrations, as well as the complexity of the photorefractive effect.

\section{Experimental set-up}

\begin{figure}[htbp]
\begin{center}
\includegraphics[width=9cm]{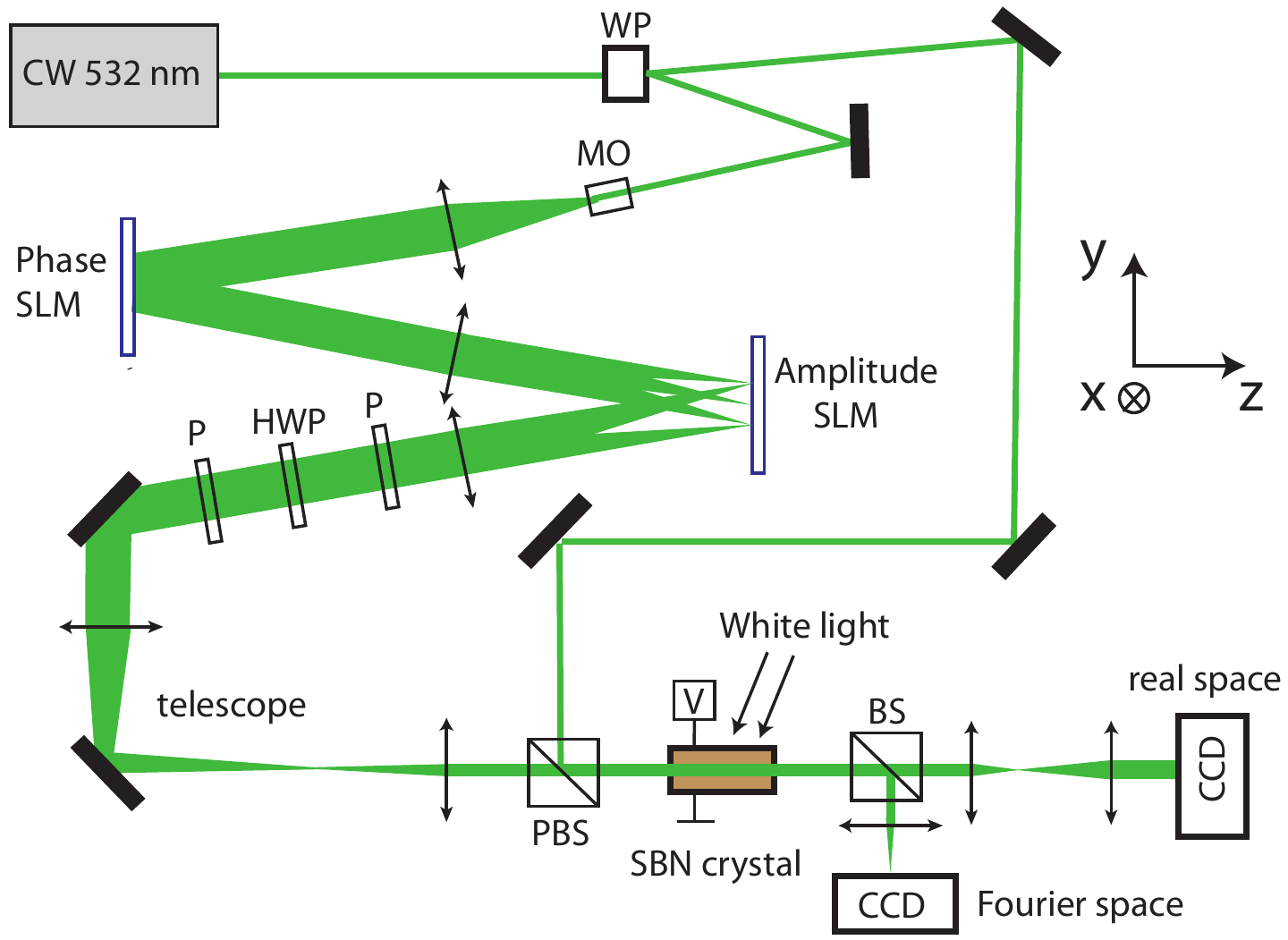}
\caption{Experimental set-up. WP : Wollaston prism. MO : microscope objective. P : polarizer. HWP : half-wave plate. PBS : polarizing beam splitter. BS : beam splitter. The lattice writing beam is modulated with a phase SLM in real space then in Fourier space with an amplitude SLM. The probe beam is imaged in real and Fourier space.}
\label{fig.setup}
\end{center}
\end{figure}

For inducing photorefractive lattices and studying them, we use standard techniques, as sketched in Fig. \ref{fig.setup}. A continuous wave laser beam at wavelength $\lambda= 532nm$ is split in two components of linear polarization. The ordinary polarized beam is used as a lattice writing beam of intensity $\IW = 1.6$mW/cm$^2$, being modulated in real space with a phase SLM (Holoeye Pluto) and dynamically filtered in Fourier space using an amplitude SLM (Holoeye LCR-1080). This allows to realize clean non-diffracting lattice beams in any 2D geometry, provided that the transverse spectrum of the lattice waves is contained in a circle (see \cite{boguslawski11} for details on the phase masks).
The extraordinary polarized beam is used as a wide gaussian, plane-wave like probe beam whose intensity at crystal output face is imaged on the real space CCD camera. The Fourier space camera is used for the alignment of the lattice and probe beams.

We use a $10\times5\times2$ mm$^3$ SBN:75 crystal with bare ordinary (resp. extraordinary) index $\nor=2.36$ (resp. $\nex=2.33$) and relevant electro-optic coefficients $r_{33} = 1340$pm/V and $r_{13} = 67$pm/V (in our notation, the crystal c-axis is $y$). An external electric field $E_0=1.5$kV/cm is applied across the crystal during photorefractive writing and probing.
A white incoherent light source is used only for erasing the patterns before rewriting.
From the erasing time of lattice patterns in the dark ($\sim$1 day), we estimate the saturation (dark) intensity in our crystals $\Isat\sim1\mu$W/cm$^2$ i.e., we work at high saturation $\IW \gg \Isat$.

\section{Inapplicability of the digital holography method in situations of interest}

To calibrate the refractive index in lattices, as well as in non-periodic patterns, the method of digital holography has been proposed in \cite{zhao03}, and used in several studies (for example,  \cite{zhang04, zhang10, boguslawski12}).
This method relies on recording the dephasing $\Delta \phi(x,y)$ of a plane wave beam after linear propagation through the crystal, using the interference with a plane wave, flat-phase reference beam. The refractive index modulation $\Delta n(x,y)$ inside the crystal, assumed invariant along $z$, is then obtained as
\begin{equation}
\Delta n(x,y) = \Delta \phi(x,y)/k L,
\label{eq.digholo}
\end{equation}
where $L$ is the crystal length and $k= 2 \pi/\lambda$ the wave vector of the probe wave.

Equation (\ref{eq.digholo}) shows that digital holography relies on the assumption, generally not fulfilled in situations of interest, that light rays propagate in the crystal along rectilinear paths of constant $\Delta n$, or in other words, that the  whole crystal behaves as a pure (thin) phase mask.
However, when the propagation distance is larger than the characteristic diffraction length \cite{kivshar03}
\begin{equation}
\ld =n_0 d^2/\lambda, 
\label{eq.lT}
\end{equation}
where $d$ is the lattice period and $n_0$ the crystal refractive index, phase modulations will modify the initially rectilinear propagation of the plane wave, light rays will bend, and the beam will develop longitudinal quasi-oscillations (LQO) of the phase and intensity, somehow analog to the Talbot effect \cite{berry96, iwanow05}. 

Thus, Eq. (\ref{eq.digholo}) is valid only for thin lattices of length $ L \ll \ld$. For our parameters, this makes digital holography applicable only for lattices with large period  
$d\gg \sqrt{L \lambda/n_0} \simeq 48\mu \rm{m}$, i.e., for very slowly varying refractive index patterns (or even flat patterns, as in \cite{bekker98}). 
As such, digital holography cannot be applied for most waveguide arrays (photonic crystals) since those structures are precisely expected to guide the light, i.e. affect strongly its intensity distribution, and interesting physical phenomena require $ L \gg \ld$.

\begin{figure}[htbp]
\begin{center}
\includegraphics[width=11.6cm]{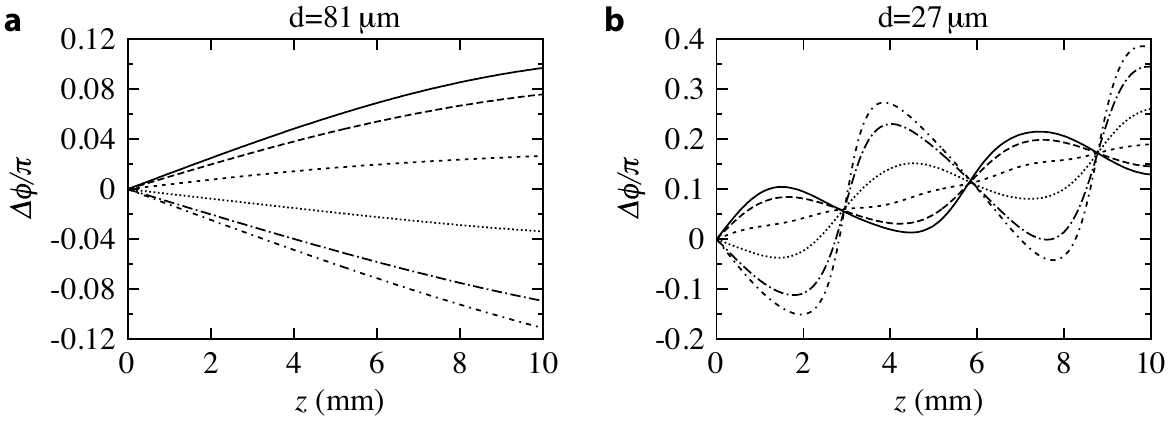}
\caption{Phase evolution of a plane wave probe beam at six regularly spaced positions $y/d=0, 0.1, 0.2, 0.3, 0.4, 0.5$ (respectively shown with lines from solid to dot-dashed) in a sinusoidal 1D lattice.
(a) Lattice period $d=81\mu$m and $\Delta n_0=0.67\times10^{-5}$.
(b) $d=27\mu$m and $\Delta n_0=0.6\times10^{-4}$. 
}
\label{fig.phi}
\end{center}
\end{figure}

To illustrate the failure of Eq. (\ref{eq.digholo}) and digital holography, Fig. \ref{fig.phi} shows the evolution of the phase $\Delta \phi(y,z)$ at six regularly spaced locations in a 1D lattice (see Eq. (\ref{eq.deltan})), in the linear propagation regime (Eq. (\ref{eq.se}))
for two lattice periods $d$ and lattice depths $\Delta n_0$ such that the product $d^2 \times \Delta n_0$ is constant (ensuring equal waveguiding strength, according to Eq. (\ref{eq.deeplatt})).
In Fig.Ê\ref{fig.phi}(a), $d=81\mu$m and Eq. (\ref{eq.digholo}) is approximately valid, as expected.
In Fig.Ê\ref{fig.phi}(b), with a more realistic $d=27\mu$m, $\Delta \phi$ is clearly not proportional to the local $\Delta n$ and to $z$, as in Eq. (\ref{eq.digholo}). In particular, one can note the periodic reconstruction of a flat phase.
Thus, digital holography cannot be applied here.
For example, in \cite{zhang10}, refractive index distributions estimated using digital holography are plotted without vertical axis, and compared only qualitatively to simulations. In this work the lattice period was $17\mu$m or less, and the crystal length $L=10$mm. Our analysis shows that the results of digital holography there are very unlikely correct quantitatively.

\section{Our method for absolute calibration of lattices}

Our method is based on analyzing the amplitude of spatial intensity modulation acquired by a plane wave probe beam during propagation through a lattice.
This approach is inspired by the method for calibrating optical lattices for Bose-Einstein condensates (BEC) \cite{denschlag02}, where the sudden turn on and off of the optical lattice is analogous in our case to the sudden entrance and outcoupling of the light wave in the photonic crystals.
It is also related to the "waveguiding technique", which is commonly used (e.g. in \cite{terhalle06, desyatnikov06}) to visualize the refract index structure, although, in those works, the refractive index was not quantitatively determined.

\subsection{Principle}

Let us first describe the ideal propagation of a plane wave in a periodic potential.
In the paraxial approximation, the propagation along $z$ of a wave of amplitude $\Psi(x,y,z)$ along a medium with a transverse refractive index $\Delta n (x,y)$ obeys a (2+1)D Schr\"odinger type equation \cite{kivshar03}
\begin{equation}
i \frac{\partial \Psi}{\partial z} = -\frac{1}{2 \beta_0} \gradp \Psi - \frac{\beta_0}{n_0} \Delta n (x, y) \Psi ,
\label{eq.se}
\end{equation}
where $n_0=\nex$, $\beta_0=2\pi n_0/\lambda$ is the propagation constant in the crystal, $\gradp = \left(\frac{\partial ^2}{\partial x ^2} + \frac{\partial ^2}{\partial y ^2} \right)$ is the transverse laplacian operator, the longitudinal coordinate $z \leftrightarrow t $ plays the role of time $t$, and where the potential $V(x,y)$ is here replaced by the refractive index : $V(x,y) \leftrightarrow -\Delta n(x,y) $.

For 1D lattices, we model the refractive index as a sinusoidal in the c-axis direction $y$ :
\begin{equation}
\Delta n (y)= \Delta n_0  \sin^2(\kL y/2 + \phi_L),
\label{eq.deltan}
\end{equation}
where $\kL=\pi/d$, with $d$ the lattice period. 
Although we use writing intensities $\IW\gg\Isat$, the sinusoidal assumption may be not bad as long as we consider the transient regime of photorefractive writing, where writing speeds are proportional to local intensities \cite{allio14}, so that patterns are not strongly saturated. For measurements at high saturation beyond the transient regime, the sinusoidal ansatz may cause larger imprecision.
We assume a lattice invariant in the $z$ direction, i.e., we implicitly neglect the absorption \cite{alpha}, as well as any non-linear effects in the writing beam.
The latter approximation relies on the strong anisotropy of the electro-optic coefficients. We however note, in the following, that lattice imperfections are strong enough to dramatically damp the longitudinal oscillations, and also cause significant aberrations for 2D lattices.

Our goal is to determine $\Delta n_0$, for any writing time $\tW$ during which the writing beam has been applied.
To do so, we send a wide, plane-wave like probe beam at normal incidence through the crystal, whose intensity at the crystal output is spatially modulated with the periodicity of the lattice. 
We then fit the vertically integrated profile with a function
\begin{equation}
I(y) = I_0 \big[1 + \alpha_1 \cos(\kL y + \phi_1) \big],
\label{eq.fit}
\end{equation}
where $I_0$ is the average beam intensity.
The coefficient $\alpha_1$ is analogous to the Fourier space amplitude of Bragg diffraction into the first orders ($\pm 2 \kL$) (see \cite{terhalle07, apolinar08, denschlag02}), noting however that by construction, $\alpha_1$ can exceed 1.
Using real space images allows us to monitor more directly the phenomena, including the appearance of parasitic (e.g. residual nonlinear) effects.

For deep lattices, where the potential energy overcomes the diffraction (kinetic energy) term in Eq. (\ref{eq.se}), i.e., satisfying the condition
\begin{equation}
\Delta n_0 \gg \lambda^2 / (2 n_0 d^2) ,
\label{eq.deeplatt}
\end{equation}
the probe light is more strongly modulated, with $\alpha_1\sim 1$, and higher harmonics become evident in the profiles.
In our data, $\alpha_1$ saturates to a value of about 1.1 for strong lattices (see, e.g. Figs. \ref{fig.damping}(a)--\ref{fig.damping}(d) or Fig. \ref{fig.1Dcalib}(d)), thus no more information is contained in its value.
In this case, we extend the analysis to harmonics up to third order, using the fitting function
\begin{equation}
I(y) = I_0 \big[1  + \alpha_1 \cos(\kL y + \phi_1) + \alpha_2  \cos(2 \kL y + \phi_2)  + \alpha_3 \cos(3 \kL y + \phi_3)   \big]
\label{eq.fit2}
\end{equation}
where the modulation coefficients $\alpha_1, \alpha_2, \alpha_3$, analogous to diffraction amplitudes of various orders in Fourier space \cite{apolinar08}, now contain the information about the lattice strength.

\subsection{Longitudinal quasi-oscillations versus Talbot effect}

\begin{figure}[htbp]
\begin{center}
\includegraphics[width=12.2cm]{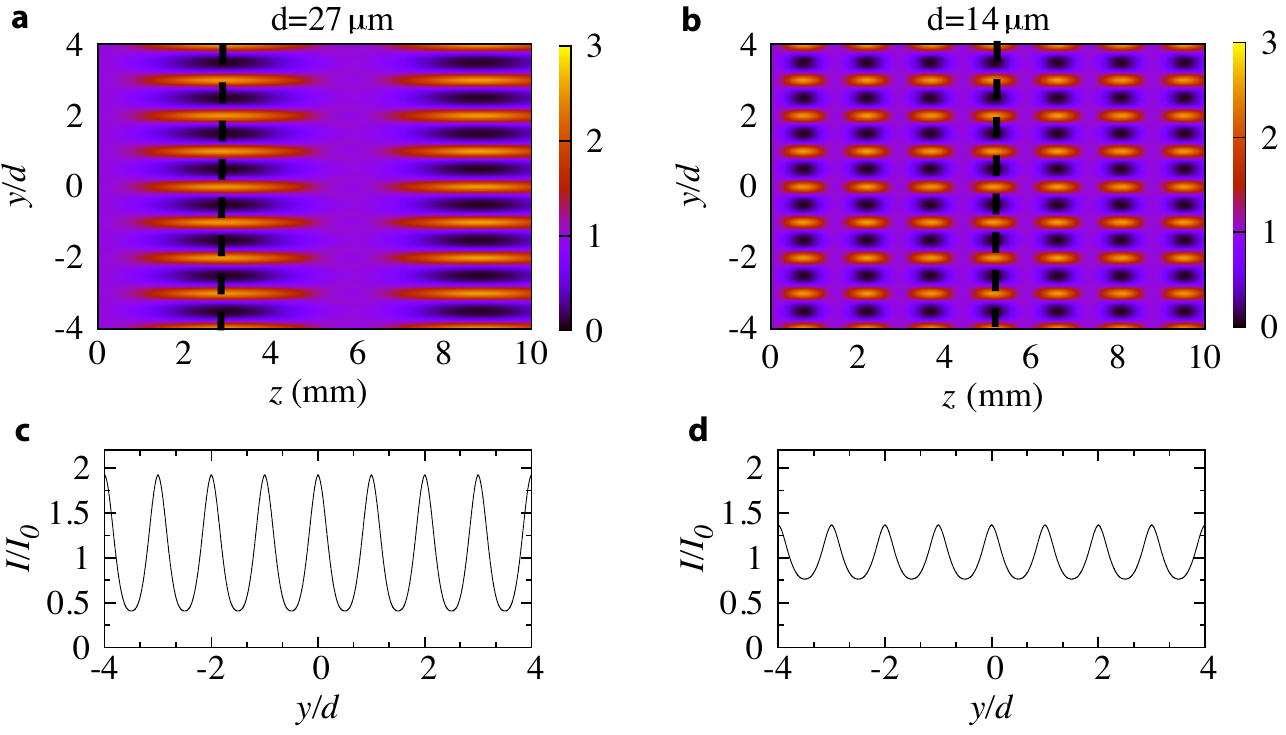}
\caption{Simulation using Eq. (\ref{eq.se}) of the ideal propagation of a plane wave in a 1D sinusoidal lattice, without longitudinal damping.
(a,b) Intensity distributions along crystal length $L=10$mm for 8 lattice periods.
(c,d) Intensity profiles at the output face $z=L$.
(a,c) Lattice period $d=27\mu$m and $\Delta n_0 = 0.6 \times 10^{-4}$. 
(b,d) $d=14\mu$m and $\Delta n_0 = 2.4 \times 10^{-4}$.
Vertical dashed lines show positions of maximally modulated profiles $\Imax(y)$.}
\label{fig.simul}
\end{center}
\end{figure}

In Fig. \ref{fig.simul}, we show simulations results for the propagation of a plane wave injected in the 1D model lattice potential of Eq. (\ref{eq.deltan}).
Figures \ref{fig.simul}(a) and \ref{fig.simul}(b) show the simulated two-dimensional intensity distributions of light in the transverse and longitudinal directions, for lattices of period $d=27\mu$m and $14\mu$m, having equal waveguiding strength (i.e., according to Eq. (\ref{eq.deeplatt}), equal value of $d^2 \times \Delta n_0$).
Figures \ref{fig.simul}(c) and \ref{fig.simul}(d) show the transverse intensity profiles at crystal output. 
In Figs. \ref{fig.simul}(a) and \ref{fig.simul}(b), the amplitude of modulation of the probe beam displays high contrast oscillations during propagation along $z$, closely connected to the phase oscillations mentioned in Section 3. 

The longitudinal quasi-oscillations (LQO) and periodic revivals are analog to the Talbot effect \cite{berry96, iwanow05}, but different in nature.
Originally the Talbot effect described a wave propagating in free space beyond a diffraction grating. Then, the field at the grating is periodically reconstructed in revivals occurring exactly at even integer multiples of the Talbot length $\zT=d^2 / \lambda$. This occurs for any periodic grating. Additionally, fractional self images are obtained at rational fractions of the Talbot period, and fractal images elsewhere \cite{berry96}.

On the other hand, the LQO inside a photonic crystal occur inside the lattice potential. 
Since they are governed by a Schrodinger equation, one can understand them in complete analogy to the case of the sudden loading of BECs in optical lattices, and thus simply adapt the description of \cite{denschlag02}. 
When the plane wave with transverse momentum $k=0$ enters the lattice, it is suddenly not anymore an eigenstate of the free space hamiltonian. 
The lattice eigenstates with quasi-momentum $k=0$ are Floquet-Bloch waves $\chin = \sum_m  a_{nm} \phim$, written here in the basis of the plane waves $\phi _m (y) = e^{2im \kL y}$ with the appropriate momenta.
At the crystal input face, the state is $\vert \Psi(0) \rangle = \sum_n a^*_{n 0} \chin$. Then, at each position $z$ along the crystal the state is
\begin{equation}
\vert \Psi(z) \rangle = \sum_n a^*_{n 0} e^{i \beta_n z}  \chin= \sum_{n, m} a^*_{n 0} a_{n m} e^{i \beta_n z} \phim,
\label{eq.psiz}
\end{equation}
i.e. a sum of plane waves that acquire different phases according to the propagation constants $\beta_n$ of the different Bloch waves.
The LQO result from the interference between the terms in Eq. (\ref{eq.psiz}), which depends on the energy differences between the Bloch waves and thus on the lattice strength. For shallow lattices only the two lowest bands are involved, thus the interference pattern involves only one frequency, and the LQO are almost truly periodic.
For strong lattices, the initial wave gets diffracted into several orders (i.e., several higher bands), and the resulting interference is not in general periodic (hence the name "quasi-oscillations"), since the $\beta_n$ have no reason to be commensurate with each other. In contrast, Talbot revivals are exceptionally robust and structured because they happen in vacuum, where all the plane waves $\phim$ involved in the decomposition of any initial periodic image (the grating) have integer multiple frequencies.

\begin{figure}[htbp]
\begin{center}
\includegraphics[width=5.5cm]{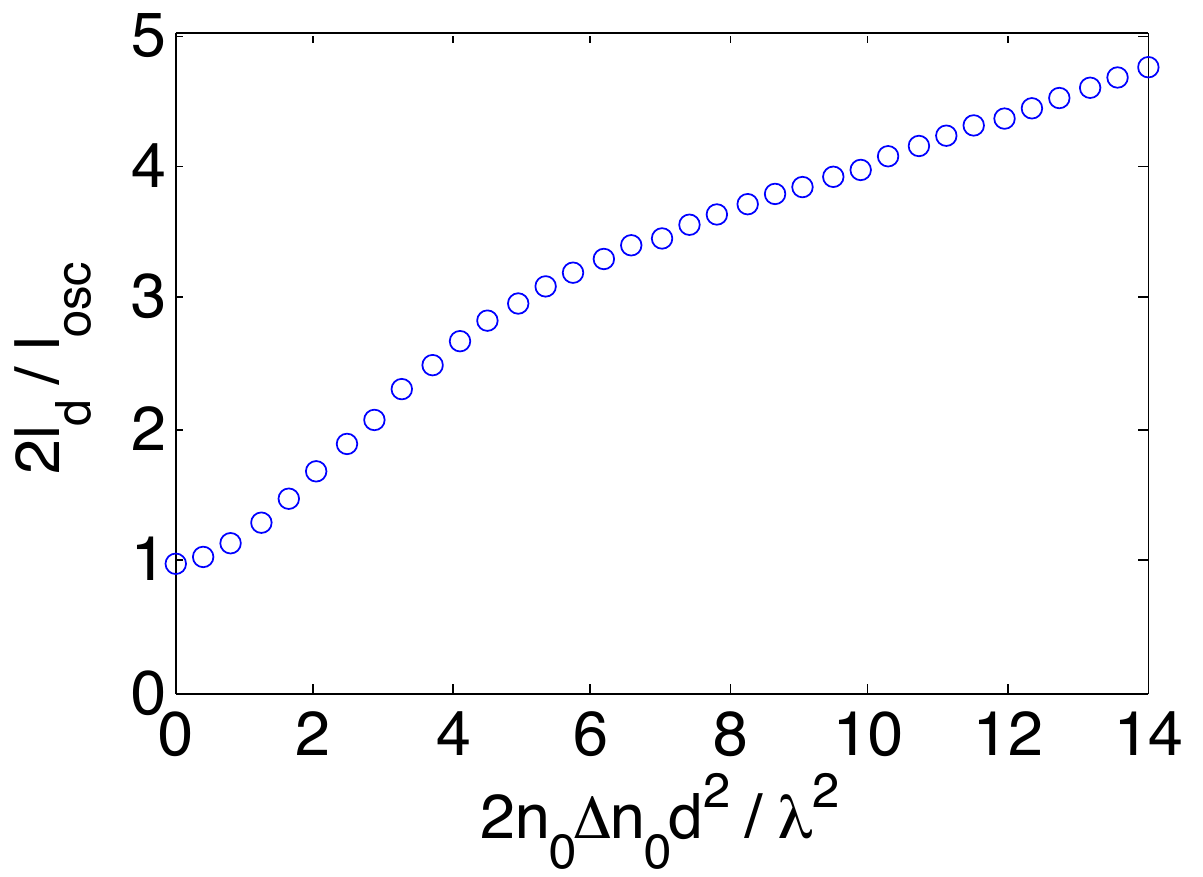}
\caption{Inverse quasi-period  $\losc$ of the longitudinal oscillations in units of $2 \ld$ as function of lattice strength $\Delta n_0$ adimensionalized according to Eq. (\ref{eq.deeplatt}), for sinusoidal 1D lattices. $\losc$ is obtained with a cosinusoidal fit of the central intensity I(y=0,z) in simulations.}
\label{fig.freq}
\end{center}
\end{figure}

Another important difference between Talbot revivals and LQO in photonic lattices is that the quasi-period $\losc$ of the LQO depends on the lattice eigenenergies and thus the lattice strength (or inter-site coupling in \cite{iwanow05}). 
Consequently, measuring this quasi-frequency provides a reliable calibration of the lattice strength in optical lattices \cite{denschlag02}.
In Fig. \ref{fig.freq}, we plot $\losc$ for a 1D sinusoidal lattice as function of the lattice strength $\Delta n_0$.
For shallow lattices (small $\Delta n_0$), the lattice eigenstates are very close to free space plane waves, and $\losc$ coincides with $2 \ld$ which is the Talbot period in a medium of index $n_0$.
In the case of Figs. \ref{fig.phi}(b) and \ref{fig.simul}(a), where $2 n_0 \Delta n_0 d^2 \lambda^2 = 0.72$, the longitudinal quasi-period is indeed slightly smaller than $2 \ld=6.4$mm.
For stronger lattices, contrary to Talbot oscillations, $1/\losc$ becomes larger while the LQO lose periodicity. 
True Talbot behavior is retrieved in the shallow lattice (free space) limit.


\subsection{Observed damping of the longitudinal oscillations}

\begin{figure}[htbp]
\begin{center}
\includegraphics[width=12cm]{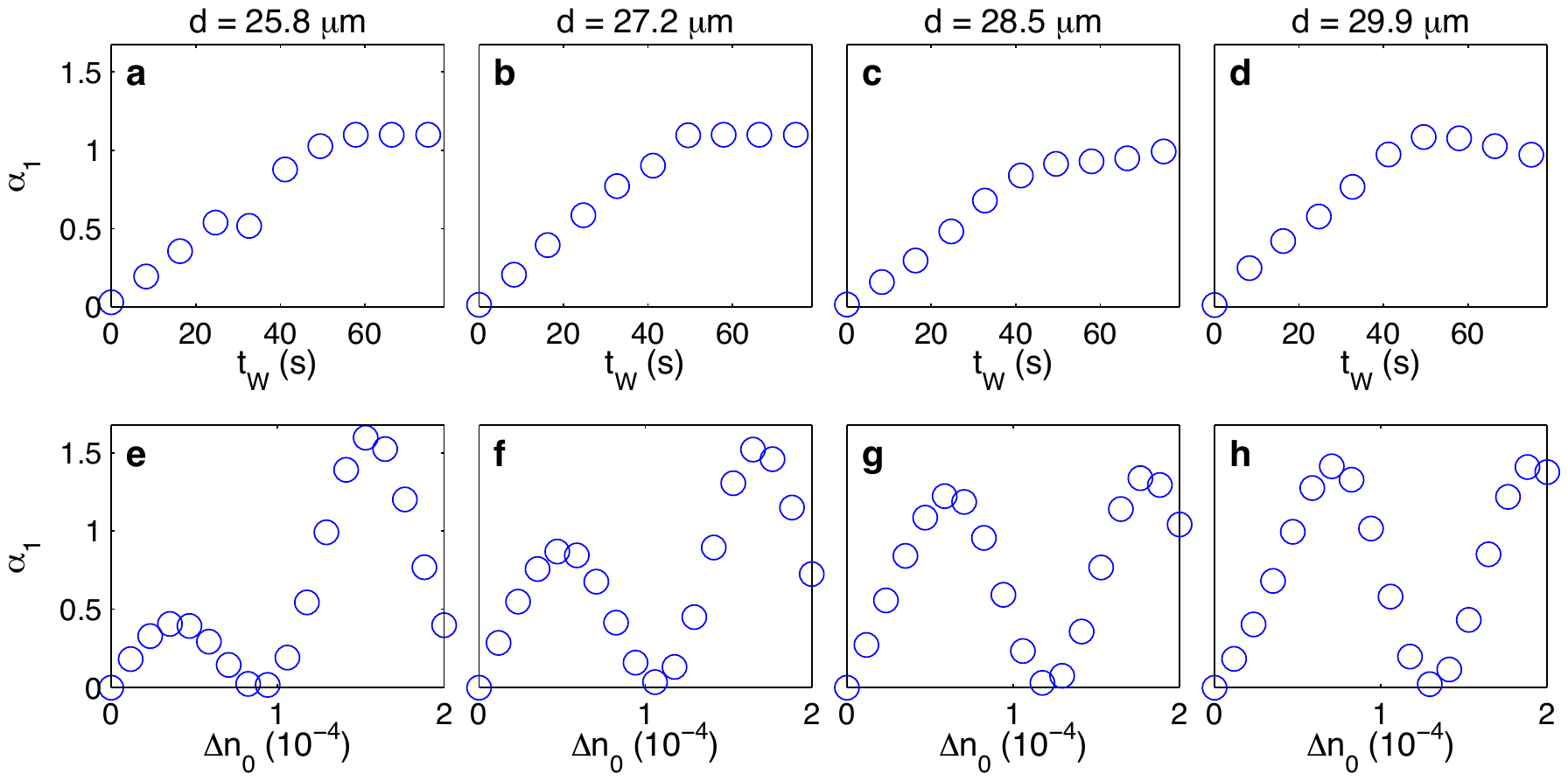}
\caption{Experimental evidence for a strong damping of longitudinal oscillations of modulation.
(a-d) Measured modulation coefficient $\alpha_1$ at crystal output as function of writing time $\tW$ for four slightly different lattice periods $d$.
(e-h) Theoretically expected $\alpha_!$ at crystal output for increasing lattice amplitudes $\Delta n_0$, assuming ideal propagation.}
\label{fig.damping}
\end{center}
\end{figure}

In our system, we can unfortunately not measure the quasi-period $\losc$ and use the calibration method for BECs in optical lattices \cite{denschlag02} since our observation plane is only the crystal output face and thus we do not have access to the time (longitudinal) evolution.
However, in analogy with \cite{iwanow05}, the LQO, if present, should be observable at the fixed output face position, by varying the lattice strength $\Delta n_0$ and thus the quasi-period (see Figs. \ref{fig.damping}(f)--\ref{fig.damping}(h).
It turns out that the LQO, expected in the ideal model, are absent, or at least, strongly damped at the observation plane.

In Fig. \ref{fig.damping}, we show the measured output modulation coefficients $\alpha_1$ for 1D lattices of four slightly different lattice periods $d=25.8, 27.2, 28.5, 29.9 \mu$m (a-d), as well as the ideal theory predictions (e-h), for a relevant range of values of $\Delta n_0$. The measured $\alpha_1$ evolve smoothly and almost identically for the four periods, whereas the theory shows very different and strongly non-monotonic behaviors such that very different and non-monotonic evolutions for $\alpha_1$ are expected for the different lattice periods $d$. This is because any change in $d$ changes the quasi-frequency of oscillations (see Fig. (\ref{fig.freq})), so that the observation plane may coincide with either a minimum or a maximum of modulation (see Fig. (\ref{fig.simul})).
From Fig. \ref{fig.damping}, we conclude that the LQO are strongly damped in the experiment \cite{data1}.

More precisely, the LQO die out sufficiently rapidly that no clear sign of them is visible at the crystal output. As for BECs in optical lattices, where important damping is observed already for 8 periods \cite{denschlag02}, we can attribute the damping to lattice imperfections, which cause loss of coherence of the different Bloch waves and blur the resulting interference. 
In photorefractive lattices, it is easy to find candidates of parasitic effects and imperfections possibly causing the observed fast damping (over less than 10 periods), for example, the residual nonlinearity in the writing beam, or diffusive mechanism (see, e.g., \cite{allio14}).

\subsection{Phenomenological model}

To calibrate our lattices, we need a model to interpret the measured profiles.
For this, we construct a simple phenomenological model. This model is heuristic and does not rely on a microscopic modeling of the lattice imperfections that cause damping of oscillations, whose detailed modeling seems very complex and lies clearly beyond the scope of this work.

Our model is mathematically the simplest that we found able to fulfill the two requirements that : 

(a) Longitudinal oscillations should be damped at the crystal output

(b) The value of $\alpha_1$ should saturate to 1.1 for strong lattices, as observed in all our data.

Starting from the previous ideal model, the simplest procedure to remove the oscillations is to simply average the profiles over several quasi-periods to construct an effective profile $\Iavg(y)$ at crystal output
However, we notice that with this procedure only, $\alpha_1$ saturates to about 0.7, i.e., notably less than the observed value of 1.1.

To understand why the observed values reach 1.1, a natural hypothesis is that the LQO are not only damped, but also, that they relax towards the ground state of the lattice, which, for deep lattices, is very close to a series of gaussians localized at each lattice site.
In this case however, $\alpha_1$ can reach values well larger than 1.1.
In the reported damped oscillations for BECs in optical lattices \cite{denschlag02}, one also observes that the first order diffraction amplitude not only displays damped oscillations, but also that these oscillations drift towards an increasing average value larger than 0.5. This is very analogous to our observation.

To account for this effect, we -again, heuristically- consider, during the propagation along $z$ in the ideal model, the maximally modulated profile $\Imax(y)$, shown for illustration in Fig. \ref{fig.simul}(a) and \ref{fig.simul}(b) as dashed vertical lines.
Finally, to match requirements (a) and (b), we construct an effective intensity profile as a weighted average 
\begin{equation}
\Ieff(y) = \eta \Iavg(y) + (1 - \eta) \Imax(y),
\label{eq.Imodel}
\end{equation}
where both $\Iavg(y)$ and $\Imax(y)$ are determined numerically from the simulations of the ideal propagation model.
The effective model Eq. (\ref{eq.Imodel}) has only one parameter $\eta=0.6$, that is easily determined from the condition of matching the saturation value of $\alpha_1= 1.1$.

Our construction of $\Ieff(y)$ is clearly phenomenological, it lacks microscopical grounding, and could perhaps be improved.
Nevertheless, it seems sufficient for our purpose of obtaining rough but absolute and consistent estimates of refractive indexes in our photo-induced lattices.

\section{Absolute calibration of 1D lattices}

Our calibration method consists in first measuring, for any writing time $\tW$, the modulation coefficients $\alpha$. Independently, we compute the theoretical values expected from the effective profiles $\Ieff(y)$, for different lattice strengths $\Delta n_0$. The calibration is performed, for each time $\tW$ independently, by numerically finding the lattice strength $\Delta n_0$ that best reproduces the experimental values of the $\alpha$ coefficients (minimizing the r.m.s. error) \cite{resol}.

\begin{figure}[htbp]
\begin{center}
\includegraphics[width=13.5cm]{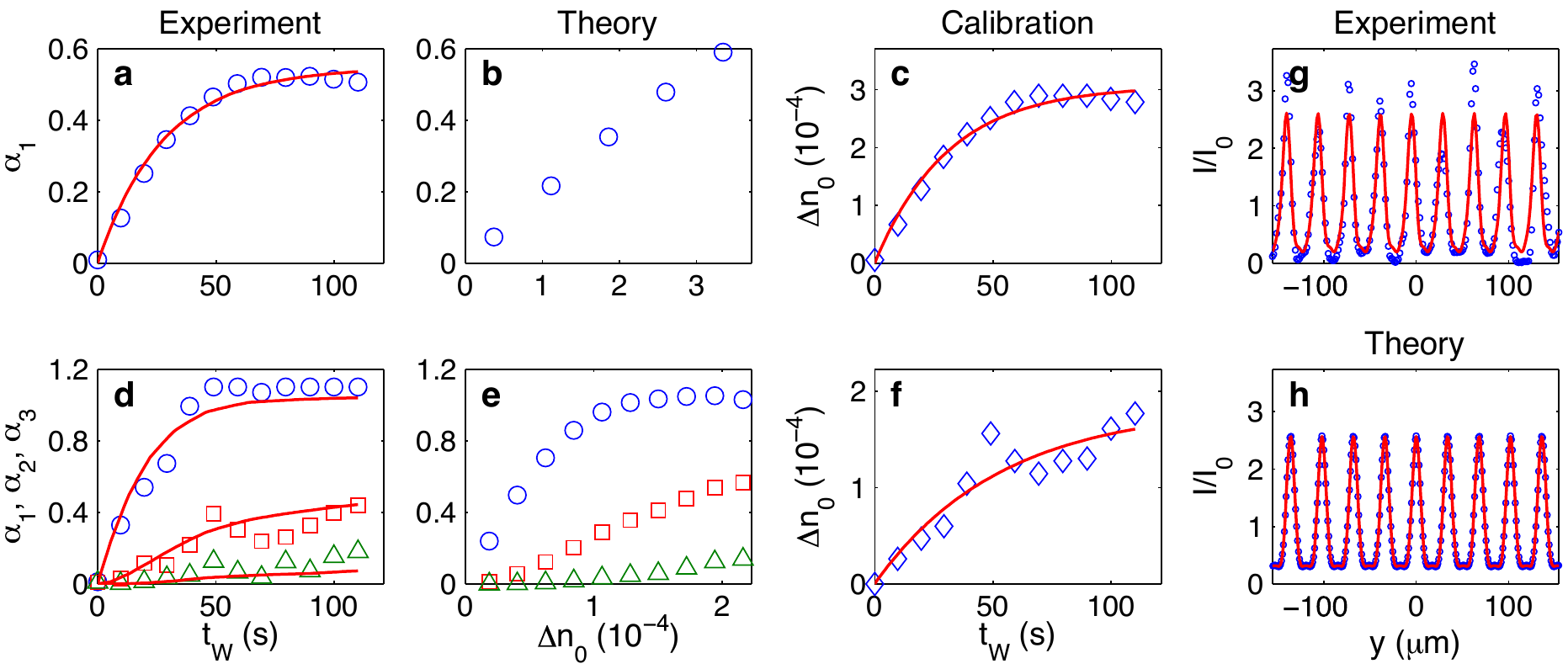}
\caption{Absolute calibration of the refractive index for 1D lattices of period $d=14\mu$m (a-c) and $d=34\mu$m (d-f). 
(a,d) Measured modulation coefficients $\alpha_1$ (circles), $\alpha_2$ (squares), $\alpha_3$ (triangles), vs writing time $\tW$. 
(b,e) Same coefficients in the effective model, as function of $\Delta n_0$.
(c,g) The resulting calibrations of $\Delta n_0$ as function of $\tW$.
The solid line is a smoothing fit using Eq. (\ref{eq.expfit}), from which we obtain the solid lines in (a,d).
(g) Measured profile for $d=34\mu$m (dots). The fit with Eq. (\ref{eq.fit2}) (solid line), yields $\alpha_1=1.10, \alpha_2=0.40, \alpha_3=0.12$. 
(h) The corresponding theoretical profile (dots) with $\Delta n_0 = 1.48\times 10^{-4}$, and $\alpha_1=1.05, \alpha_2=0.44, \alpha_3=0.07$.}
\label{fig.1Dcalib}
\end{center}
\end{figure}

Figure \ref{fig.1Dcalib} presents the absolute calibration of 1D lattices of periods $d=14\mu$m (Figs. \ref{fig.1Dcalib}(a)--\ref{fig.1Dcalib}(c)) and $d=34\mu$m (Figs. \ref{fig.1Dcalib}(d)--\ref{fig.1Dcalib}(f)), for different writing times $\tW$.
For the $d=14\mu$m lattice, $\alpha_1$ is sufficient.
Figure \ref{fig.1Dcalib}(a) shows the measured $\alpha_1$ at different $\tW$, Fig. \ref{fig.1Dcalib}(b), the predicted $\alpha_1$ for different values of $\Delta n_0$ using our effective theory, and Fig. \ref{fig.1Dcalib}(c) , the resulting values of $\Delta n_0$ estimated independently for each $\tW$.
Figures \ref{fig.1Dcalib}(d)--\ref{fig.1Dcalib}(f) show the same procedure carried for the $d=34\mu$m lattice, but using coefficients up to third order ($\alpha_1, \alpha_2, \alpha_3$).
This is useful since for larger period $d$, the deep lattice criterium of Eq. (\ref{eq.deeplatt}) is more easily reached and thus modulation coefficients are stronger, in particular $\alpha_1$ saturates to about 1.1 already at $\tW=50s$.

In the final calibrations, shown in \ref{fig.1Dcalib}(c) and \ref{fig.1Dcalib}(f), we notice a stronger noise for the $34\mu$m lattice, which is a deeper lattice according to Eq. (\ref{eq.deeplatt}).
To obtain smooth time evolutions for  $\Delta n_0$, we fit the results with exponential functions (solid lines) 
\begin{equation}
\Delta n_0(\tW) = \Delta n_0^\infty [1 - \exp(\tW/\tauW)].
\label{eq.expfit}
\end{equation}
To compare a posteriori the theory and measurements, we reconstruct, from these fits, smooth behaviors for the $\alpha$ coefficients (solid lines in Figs. \ref{fig.1Dcalib}(a) and  \ref{fig.1Dcalib}(d). 
In both cases the agreement between theory and measurements is quite satisfactory within the intrinsic noise of the data.

\section{Absolute calibration of square 2D lattices}

It is relatively straightforward to apply the same methodology to 2D lattices, although care needs to be taken with the lattice anisotropy.
For simplicity, we use square lattices, both in measurements and simulations, writing the index of refraction 
\begin{equation}
\Delta n (x, y)= \Delta n_0 \frac{\chi \sin^2(\kL y/2) + \sin^2(\kL x/2) }{1 + \chi},
\label{eq.deltan2d}
\end{equation}
where $\chi$ quantifies the anisotropy of the lattice.
This form is adpated to 2D photorefractive lattices, where the lattice period is the same in the strong direction $y$ (c-axis) and the weak direction $x$, but with an anisotropic amplitude of modulation. Note however that this representation is quite simplified compared to the expected form of the refractive index in square lattice \cite{desyatnikov06}. For discussions of the photorefractive anisotropy, see, e.g. \cite{zozulya95, allio14, terhalle07}.

\begin{figure}[htbp]
\begin{center}
\includegraphics[width=13.5cm]{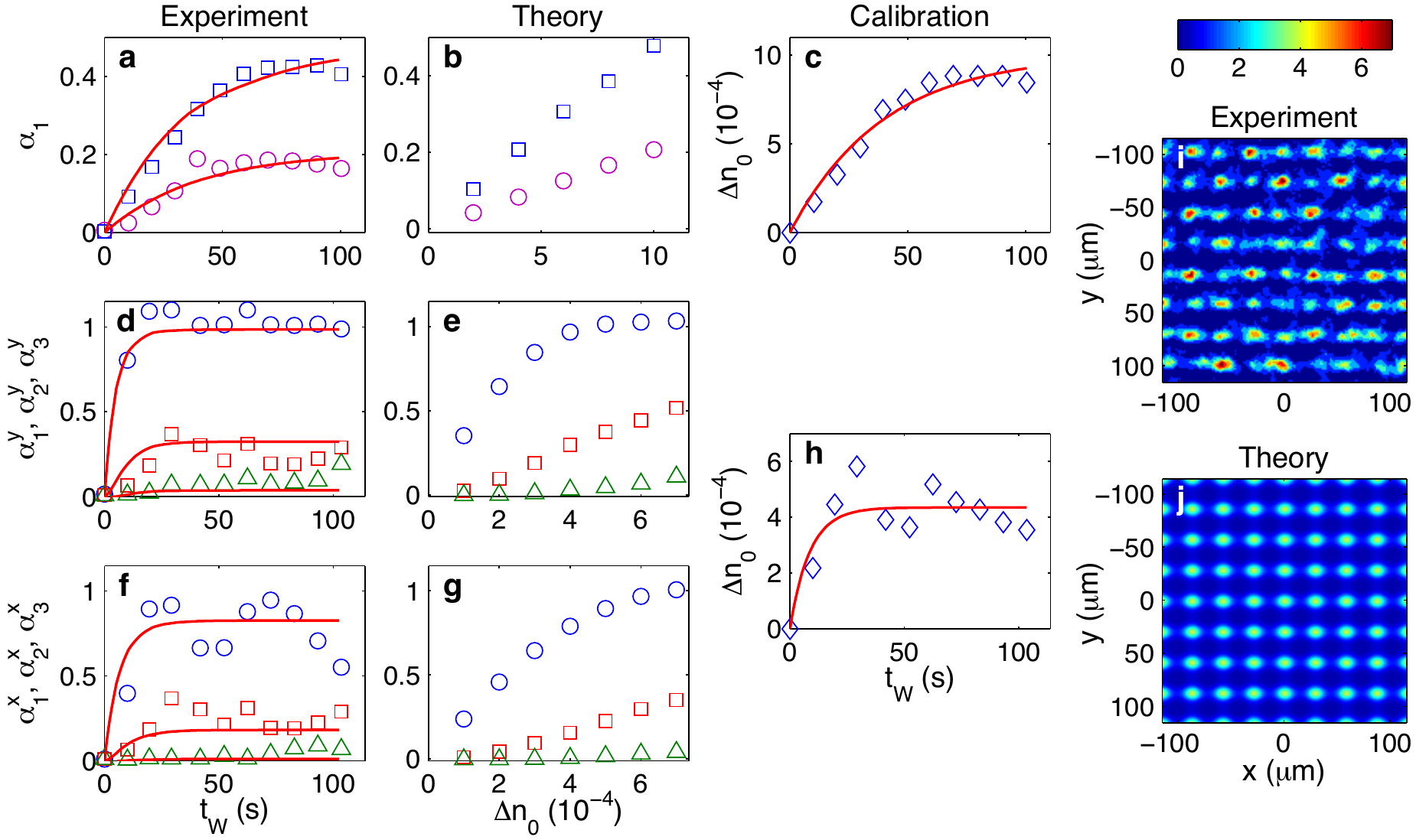}
\caption{Absolute calibration of refractive index for 2D lattices of period $d=14\mu$m (a-c) and $d=38\mu$m (d-h).
(a) Measured  coefficients $\alpha_1^y$ (squares) and $\alpha_1^x$ (circles) vs writing time $\tW$. 
(d,f) Measured $\alpha_1$ (circles), $\alpha_2$ (squares), $\alpha_3$ (triangles) in $y$ and $x$ directions.
(b,e, g) Same quantities in the effective model as function of $\Delta n_0$.
(c,h) Final estimation of $\Delta n_0$ as function of $\tW$. The fitted anisotropies are $\chi=2.5$ for $d=14\mu$m and $\chi=1.5$ for $d=38\mu$m.
The solid line is a fit with Eq. (\ref{eq.expfit}), from which we obtain the theory curves (solid lines) in (a,d,f).
(i) Experimental output intensity $I/I_0$ for $d=29 \mu$m, and $\tW=59$s. Measured coefficients are $\alpha_1^y=1.1$, $\alpha_2^y=0.32$, $\alpha_3^y=0.05$; 
$\alpha_1^x=0.36$, $\alpha_2^x=0.04$, $\alpha_3^y=0.01$.
(j) The corresponding intenstity distribution in the effective model (same color scale), for $\Delta n_0 =6\times 10^{-4}$ and $\chi=2.0$, with $\alpha_1^y=0.96$, $\alpha_2^y=0.30$, $\alpha_3^y=0.04$; $\alpha_1^x=0.62$, $\alpha_2^x=0.09$, $\alpha_3^y=0.006$.}
\label{fig.2Dcalib}
\end{center}
\end{figure}

In our measurements, we cannot detect a variation of anisotropy as function of $\tW$, therefore, we assume that $\chi$ is constant for a given lattice period $d$, at any writing time (i.e. for any lattice depth).
For shallow lattices, the ratio $\beta=\alpha_1^y / \alpha_1^x$, where $\alpha_1^y$ and $\alpha_1^x$ are 1D modulation coefficients in the $y$ and $x$ directions, is simply proportional to $\chi$. 
In the general case, to determine $\chi$ for each $d$, we fit our effective model with adjustable $\chi$ to the measurements. 

In Fig. \ref{fig.2Dcalib}, we show the calibration method applied to square 2D lattices of periods $d=14\mu$m  (a-c) and  $d=38\mu$m (Figs. \ref{fig.2Dcalib}(d)--\ref{fig.2Dcalib}(h)). 
The lattice anisotropy $\chi$ is determined first in an independent step.
For the $d=14\mu$m data, for which $\alpha_1^x$ and $\alpha_1^y$ are small, $\chi$ is determined simply from the ratio of their maximal values $\beta=\alpha_1^y/\alpha_1^x=2.5$. For the $d=38\mu$m lattice, we use modulation coefficients up to third order.

Once $\chi$ is found, the method to estimate the lattice strength $\Delta n_0$ for each writing time $\tW$ independently (data in Figs. \ref{fig.2Dcalib}(a), \ref{fig.2Dcalib}(d) and \ref{fig.2Dcalib}(f)) is the same as for 1D lattices, but now using twice more coefficients in the fitting procedure.
The final calibrations are displayed in Figs. \ref{fig.2Dcalib}(c) and \ref{fig.2Dcalib}(h), and are fitted with Eq. (\ref{eq.expfit}) to obtain smooth interpolating curves (solid lines), from which we obtain, again, smooth time evolution curves for the $\alpha$ coefficients (solid lines in Figs. \ref{fig.2Dcalib}(a), \ref{fig.2Dcalib}(d),\ref{fig.2Dcalib}(f)).

It is important to note the considerable noise in the experimental data especially for the beam modulation in the weak direction $x$ (Fig. \ref{fig.2Dcalib}(i)). Significant non-monotonous behavior of those coefficients indeed is observed (Fig. \ref{fig.2Dcalib}(f)). This effect is systematic, and is stronger for the lattices with the larger periods. In the direct real space pictures, the drop of $\alpha_1^x$ below its first maximal value is associated with a deformation of the shape of the guided light, with a systematic pattern towards the same direction (left direction in Fig. \ref{fig.2Dcalib}(i)). Such parasitic effect may result from a non-negligible contribution of the diffusive (vs drift) photorefractive mechanism \cite{allio14}.

\section{Summary of lattice calibrations in 1D and 2D}

\begin{figure}[htbp]
\begin{center}
\includegraphics[width=10cm]{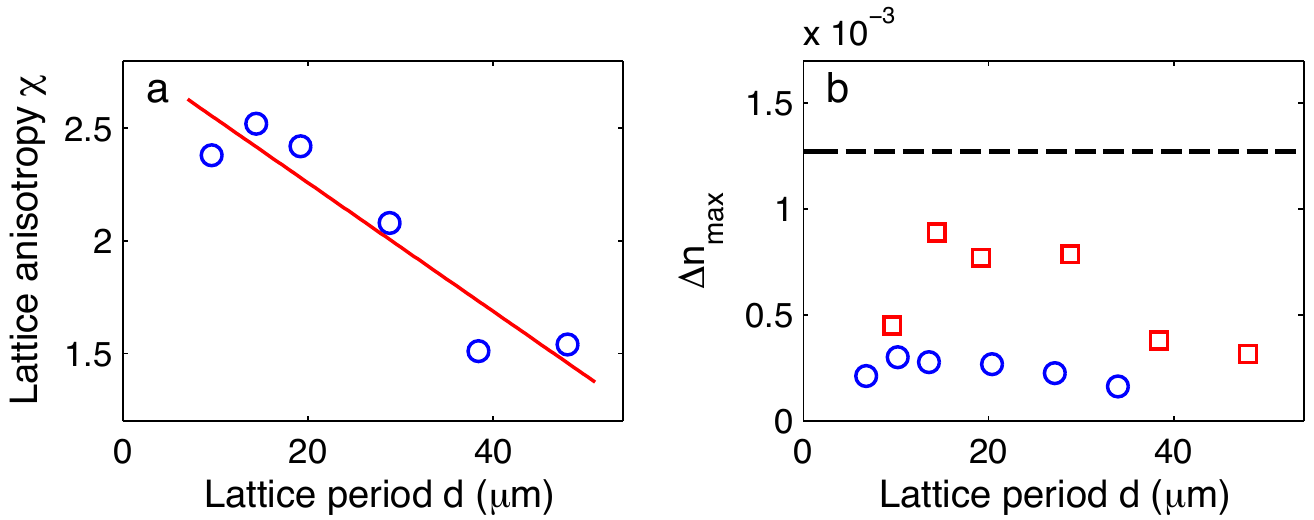}
\caption{Summary of absolute calibration of lattices with different periods $d$. (a) Lattice anisotropy $\chi$ in square 2D lattices. The solid line is a guide to the eye (linear fit to the data).
(b) Maximal refractive index $\dnmax$ for 1D lattices (circles) and square 2D lattices (squares), obtained at writing time $\tW=100$s, bias field $E_0=1.5$kV/cm, and average writing beam intensity $\IW=1.6$mW/cm$^2$. The dashed line shows the maximal value expected from standard theory in the screening regime $\dnmaxt = 0.5 n_e^3 r_{33} E_0$.}
\label{fig.summary}
\end{center}
\end{figure}

In Fig. \ref{fig.summary}, we present a summary of calibrations for lattices of periods $d=7\mu$m to $48\mu$m. 
In Fig. \ref{fig.summary}(a) , we show the lattice anisotropies $\chi$ that we find larger for smaller lattice period $d$.

In Fig. \ref{fig.summary}(b) we plot the maximal lattice depths $\dnmax$ found at $\tW=100$s in identical writing conditions, for 1D (circles) and 2D lattices (squares). At this $\tW$, for all measurements, $\dnmax$ is approaching a stationary value.
In the isotropic photorefractive theory (see e.g. \cite{fleischer03, allio14}), the steady-state refractive index modulation in the screening (drift-dominated) regime is
\begin{equation}
\dnt =  \dnmaxt \frac{1}{1  + \Isat/I},
\end{equation}
where $\dnmaxt = 0.5 n_e^3 r_{33} E_0$, independent on the lattice period.
In Fig. \ref{fig.summary}(b) , the dashed line represents this theoretical maximum for our parameters, $\dnmaxt =1.3\times 10^{-3}$.
Our measurements for 1D lattices are compatible with this theory, within a dispersion of about 30\% for the different periods $d$. 
The average value $\dnmax / \dnmaxt = 0.19$ is reasonable considering the effects of high saturation ($\IW\gg\Isat$).

For 2D lattices, the calibrations of $\dnmax$ display substantially stronger variations for different lattice periods $d$. 
In all cases, $\dnmax$ is found larger for 2D than for 1D lattices with an average $\dnmax / \dnmaxt$ A possible explanation is that in the 2D lattice the light intensity maxima are twice higher than in 1D lattices with the same average intensity, and that our measurements are still close to the transient regime where the writing speed is proportional to the local intensity (see \cite{allio14}).
For intermediate lattice periods $d=15-30\mu$m, values are higher.
As a possible explanation, one can note that parasitic effects are stronger in 2D than in 1D, as seen on Fig. \ref{fig.2Dcalib}(i), probably due, in part, to higher local intensity maxima in 2D.
Also, In their full anisotropic calculations of the refractive index induced by a gaussian beam, the authors of \cite{zozulya95} found refractive index patterns displaying, besides the local maximum, negative side lobes caused by the photorefractive anisotropy. Such lobes increase the total refractive index modulation, and may cause enhanced waveguiding that ultimately fools our method.
Contributions of the diffusive photorefactive mechanism may also be present. Finally, one should remind that the sinusoidal model that we used for the refractive index is not expected to be very accurate especially for 2D patterns and especially in high saturation and transient conditions

\section{Conclusion}
We have studied the linear propagation of plane waves in photo-induced lattices as a resource for experimentally calibrating the lattice strength using real space, near-field measurements, independently from the most often used Kukhtarev theory which relies on several hardly controllable approximations and parameters.

We first clarified theoretically the validity condition, in terms of diffraction length, that makes digital holography generally not applicable for photonic lattices.
We then found experimentally that the modulation amplitudes of a plane wave probe at the crystal output cannot be explained by an ideal propagation theory in a perfect lattice, which predicts high contrast longitudinal (Talbot-like) quasi-oscillations.
Our measurements showed, instead, that the oscillations are rapidly damped -faster than for BECs in optical lattices-, and indicated some relaxation towards the lattice ground state.

To interpret our data, we constructed a simple heuristic model accounting for these two observations, with only one parameter that is easily extracted from measurements.
Our model is unprecise and lacks of microscopic grounding, but carrying a full modeling of the lattice imperfections that cause the damping of oscillations is probably a rather unaccessible task.
Due to it simplicity, our model is robust and self-consistent.

We obtained fairly reliable direct experimental calibrations of refractive index amplitudes for several lattice periods and writing times.
For 1D lattices, our calibrations are compatible with the approximate isotropic steady-state theory.
For 2D lattices, we found anisotropies $\chi=1.5-2.5$, larger for lattices of smaller periods, and higher refractive indexes as for 1D lattices.
The measurements were systematically more noisy than in 1D, probably due to the enhancement in 2D of several non-ideal effects : parasitic nonlinearity, diffusive mechanism, anisotropy, which may cause deformation of the waveguides, for example with negative side lobes \cite{zozulya95}.
Our observations illustrate the importance of performing direct calibrations, as well as the difficulty of this task for the photorefractive system.
Possibly, in conditions of low saturation and in the stationary regime, the sinusoidal model for the refractive index may give more precise results.

As a possible complementary approach, one could consider using Fourier space data, i.e., the amplitude of diffraction peaks in the various orders \cite{apolinar08, denschlag02}.
However, the problems and solutions encountered would probably be similar to those faced in our approach, since the modulation coefficients that we use are closely connected with the diffraction amplitudes. 
It could also be interesting to revisit the methods using plane waves at the Bragg angle \cite{kukhtarev78, magnusson76, kogelnik69}, transferring them from the context of holographic recording to the context of photonic lattices.

\section*{Acknowledgments} 
We acknowledge stimulating discussions with Benjamin Pasquiou, Martin Boguslawski, Serguey Odulov, Mario Molina, and Pierre Pellat-Finet. Work supported by Programa de Financiamiento Basal de CONICYT (FB0824/2008), Pograma ICM P10-030-F and FONDECYT Grant 3140608.

%

\begin{thebibliography}{99}
\bibitem{alpha}
$\alpha=0.4$cm$^{-1}$ at $\lambda=488$nm according to fabricant (Altechna). $\alpha$ generally decreases for increasing $\lambda$, thus, at $\lambda=532$nm we expect that the absorbed fraction for our crystal is not higher than 30\%.
\bibitem{data1}
The data of Fig. \ref{fig.damping} are for lattice periods $d\simeq28\mu$m, close to our maximum. 
For smaller $d$, where longitudinal oscillations are faster, we expect even stronger damping effects.
\bibitem{resol}
We take into account the finite resolution of our imaging system by convoluting the profiles given by the effective theory by a gaussian optical response of r.m.s width $\sigma_0=0.80\mu$m.
\end{thebibliography}

\end{document}